\documentstyle[pre,preprint,aps]{revtex}
\tightenlines
\newcommand{\siml}{\stackrel{<}{\sim}}
\newcommand{\simg}{\stackrel{>}{\sim}}

\begin{document}
\draft

\title{
Responses of a Hodgkin-Huxley Neuron to
Various Types of Spike-Train Inputs
}
\author{
Hideo Hasegawa
}
\address{
Department of Physics, Tokyo Gakugei University,
Koganei, Tokyo 184, Japan
}
\date{\today}
\maketitle
\begin{abstract}

\end{abstract}
Numerical investigations have been made of
responses of a Hodgkin-Huxley (HH) 
neuron to spike-train inputs whose
interspike interval (ISI) is modulated by deterministic,
semi-deterministic (chaotic) and stochastic signals. 
As deterministic one, we adopt  inputs with the  
time-independent ISI and with time-dependent ISI
modulated by sinusoidal signal.
The R\"{o}ssler and Lorentz models are adopted for
chaotic modulations of ISI.
Stochastic ISI inputs with the Gamma distribution
are employed.
It is shown that distribution of output ISI data
depends not only on the mean of  ISIs of
spike-train inputs but also on their fluctuations. 
The distinction of  responses to the
three kinds of inputs can be made by return maps 
of input and output ISIs, but not
by their histograms.
The relation between the variations of input and 
output ISIs is shown to be different
from that of the integrate and fire (IF) model
because of the refractory period in the HH neuron.

\vspace{1.0cm}
%

\begin{center}
{\bf I. INTRODUCTION}
\end{center}

Neurons in our brain are known to be responsible 
for encoding the characteristics of stimuli into a form
for further processing by other neurons.
During the last decades the anatomical, physiological and 
theoretical studies on neurons have been extensively made.
Despite these efforts, the code used 
for encoding and decoding in neurons is
not clarified at the moment \cite{Rieke96}.
It is commonly believed that the firing rate reflects
the strength of the inputs which trigger the action
potentials of neurons. Indeed, the firing activities of 
motor and sensory neurons 
vary in response to the applied stimuli.
It is not known, however, whether the information is carried 
through the mean firing rate (rate encoding) 
or through the details of
sequences of the temporarily encoded interspike interval (ISI)
(temporal encoding), which is currently controversial
\cite{Sejnowski95}-\cite{Ferster95}.
In the last few years, experimental evidences have
accumulated, indicating  that many biological systems 
use the temporal coding. 
Human visual systems, for example, have shown to classify
patterns within 150 msec in spite of the fact 
that at least ten synaptic stages are involved from 
retina to the temporal brain \cite{Thorpe96}.
The similar speed of visual processing has been reported
for macaque monkeys \cite{Rolls94}.
Because the firing frequency of neurons involved is less than
100 Hz, each neuron can contribute
at most one or two spikes to such computations;
there is not sufficient time 
to sample firing rates.

In recent years, much studies on the encoding
of the spike trains by neurons have been made by using
the integrate and fire (IF) model \cite{Maass97}, which is one of 
the simplest, dynamical models of neurons \cite{Gerstner95}. 
The IF neutron is silent without the external, input current, 
$I_{\rm i}$.
When $I_{\rm i}$, exceeds the critical value, $I_{\rm ic}$,
the IF neuron shows the self-excited oscillations, whose
frequency, $f_{\rm o}$, depends on the magnitudes of $I_{\rm i}$.
It is shown that $f_{\rm o}$ continuously vanishes 
when $I_{\rm i}$ is decreased and approaches to $I_{\rm ic}$.
This behavior of the continuous 
$f_{\rm o}-I_{\rm i}$ dependence is different from 
the discontinuous one at $I_{\rm ic}$
in the more realistic Hodgkin-Huxley (HH) neurons \cite{Hodgkin52};
the IF and HH neurons are classified as the type I and type II,
respectively \cite{Gutkin98}.
Furthermore, the IF neuron has the disadvantages of the artificial
reset of the action potential and the lack
of the refractory time.
Although it has been widely employed for
the study of neural networks, the IF model is too crude to
discuss the activities of real neurons.

The HH model, which well describes the spiking behavior 
and refractory properties of real 
neurons, is expressed based on 
non-linear conductances of Na and K ion
channels \cite{Hodgkin52}.  
Since the HH model was proposed, its property has
been intensively  investigated 
\cite{Nemoto75}-\cite{Matsumoto84}.
The behavior of self-excited oscillations of the HH neuron
with the applied current has much variety than that
of the IF model.
It is shown that the oscillation of the HH neuron may become chaotic 
when the sinusoidal $I_{\rm i}$ is applied 
with proper choices of magnitude and frequency 
\cite{Aihara84} \cite{Matsumoto84}.
Such chaotic oscillations are experimentally observed 
in squid giant axons \cite{Matsumoto80} \cite{Guttman80}
and Onchidium neurons \cite{Hayashi85}.

The HH model was originally proposed to 
account for the property of squid giant axons \cite{Hodgkin52}
and it has been generalized with modifications 
of ion conductances \cite{Arbib95}.
The HH-type models have been widely adopted for a study
on activities of {\it transducer neurons} such as motor 
and thalamus relay neurons, which transform the 
amplitude-modulated input to spike-train outputs.
In this paper, we pay our attention to 
{\it data-processing neurons} 
which receive and emit the spike-train pulses.
Assuming that the data-processing neuron may be 
essentially described
by the ion-conductance mechanism of the HH model,
we investigate its input-output response
in order to get some insight to the following questions:

\noindent
(1) How the output ISIs depend on the input ISIs?
Does the average rate of the output ISI depend only on the
average of the input ISIs?

\noindent
(2) How do neurons distinguish the 
different types of
deterministic, chaotic and stochastic inputs?
How different is the response to different types of spike-train
inputs?

Our paper is organized as follows:
In the next sec.II, we mention a simple neuron 
model adopted for our numerical calculation.
In sec.III, we investigate the response of our system to
deterministic inputs with time-independent ISI ($\S 3.1$) and
time-dependent ISIs modulated by sinusoidal signal ($\S 3.2$).
Input and output ISIs are studied by their histograms
and return maps; the former shows the distributions
and the latter the time correlation of ISI data.
In sec.IV, chaotic inputs generated by R\"{o}ssler ($\S 4.1$)
and Lorentz model ($\S 4.2$)  are discussed.
Stochastic inputs with the Gamma distribution are
treated in sec.V.
The final section VI is devoted to conclusion and discussion

\begin{center}
{\bf II. ADOPTED MODEL}
\end{center}

We adopt a simple system consisting of a neuron and a synapse. 
The neuron is assumed to be described by the HH model and
the synapse by the alpha function (Eq.(16)).
We will investigate the response of our neuron when
spike-train inputs are applied through the synapse.

The HH model is described by the non-linear coupled differential
equations for the four variables, $V$ for the membrane potential,
and $m, h$ and $n$ for the gating variables of Na and
K channels, and it is given by \cite{Hodgkin52}
\begin{equation}
C d V/d t = -g_{\rm Na} m^3 h (V - V_{\rm Na})
- g_{\rm K} n^4 (V - V_{\rm K}) 
- g_{\rm L} (V - V_{\rm L}) + I_{\rm i},
\end{equation}
\begin{equation}
d m/d t = - (a_m + b_m) \: m + a_m,
\end{equation}
\begin{equation}
d h/d t = - (a_h + b_h) \: h + a_h,
\end{equation}
\begin{equation}
d n/d t = - (a_n + b_n) \: n + a_n,
\end{equation}
where
\begin{equation}
a_m = 0.1 \: (V + 40)/[1 - e^{-(V+40)/10}],
\end{equation}
\begin{equation}
b_m = 4 \: e^{-(V+65)/18},
\end{equation}
\begin{equation}
a_n = 0.01 \: (V + 55)/[1 - e^{-(V+55)/10}],
\end{equation}
\begin{equation}
b_n = 0.125 \: e^{-(V+65)/80},
\end{equation}
\begin{equation}
a_n = 0.07 \: e^{-(V+65)/20},
\end{equation}
\begin{equation}
b_n = 1 / [1 + e^{-(V+35)/10}].
\end{equation}
Here  the reversal potentials of Na, K channels and leakage are  
$V_{\rm Na} = 50$ mV, $V_{\rm K} = -77$ mV and 
$V_{\rm L} = -54.5 $ mV;
the maximum values of corresponding conductivities are
$g_{\rm Na} = 120 \; {\rm mS/cm}^2$, 
$g_{\rm K} = 36 \; {\rm mS/cm}^2$ and
$g_{\rm L} = 0.3 \; {\rm mS/cm}^2$; the capacity of the membrane is
$C = 1 \; \mu {\rm F/cm}^2$, details of the HH model being presented
in Refs.\cite{Hodgkin52} \cite{Park96}.

The external, input current, $I_{\rm i}$, is taken 
to consist of two terms;
\begin{equation}
I_{\rm i} = I_{\rm s} + I_{\rm p}.
\end{equation}
where $I_{\rm s}$ expresses the static dc current 
and $I_{\rm p}$ denotes the pulse
current induced by the spike-train input 
whose explicit form will be discussed shortly (Eq.(15)).

We consider the delta-function-type spike-train input expressed by
\begin{equation}
U_{\rm i}(t) = V_a \: \sum_n  \: \delta (t - t_{{\rm i}n}).
\end{equation}
The firing time $t_{{\rm i} n}$ for arbitrary $n$
is assumed to be recurrently defined by
\begin{equation}
t_{{\rm i}n+1} = t_{{\rm i}n} 
+ T_{{\rm i}n}(t_{{\rm i}n}),
\end{equation}
\begin{equation}
t_{{\rm i}1} = 0,
\end{equation}
where ISI of input spike, 
$T_{{\rm i}n}$, is
generally a function of a given time $t_{{\rm i}n}$.  
In this study, we take $T_{{\rm i}n}$ to be constant ISI,
and time-dependent ISI modulated by sinusoidal,
chaotic and stochastic signals.

The spike train given by Eq.(12) is assumed to be injected
through the synapse, yielding the current $I_{\rm p}$ given by
\begin{equation}
I_{\rm p}(t) = g_{\rm syn} \: 
\sum_n \: \alpha (t - t_{{\rm i}n}) \: (V_a - V_{\rm syn}). 
\end{equation}
Here $g_{\rm syn}$ and $V_{\rm syn}$ are the conductivity and
potential of synapse,
and the alpha function, $\alpha(t)$, 
is defined by \cite{Park96}
\begin{equation}
\alpha(t) = (t/\tau) \; e^{-t/\tau} \:  \Theta(t),
\end{equation}
where $\tau$ is the time constant relevant to the synapse conduction
and $\Theta (t)$ is the Heaviside step function.
When the ISI is very large compared with $\tau$, 
Eqs.(15) and (16) yield pulse currents
with the maximum value of 
$I_{\rm p}^{\rm max} = e^{-1} \: g_{\rm syn} \: 
(V_{\rm a} - V_{\rm syn})$
at $t = t_{{\rm i}n} + \tau$ and with the half-width
of $2.45 \: \tau$.
We assume $V_a = 30$ mV (the typical value of the
maximum membrane potential), $V_{\rm syn} = - 50$ mV and
$\tau = 2$ msec,
and treat $g_{\rm syn}$ as a parameter.

When the membrane potential $V$ oscillates,
it yields the spike-train output, which may be expressed by
\begin{equation}
U_{\rm o}(t) = V_a \: \sum_m \: \delta (t - t_{{\rm o}m}),
\end{equation}
in a way similar to Eq.(12), and  the output ISI is defined by
\begin{equation}
T_{{\rm o}m} = t_{{\rm o}m+1} - t_{{\rm o}m}.
\end{equation}
We will investigate how $T_{{\rm o}m}$ depends on 
the various types of $T_{{\rm i}n}$.

Differential equations given by Eqs.(1)-(10) including
the external current given by Eqs.(11)-(16) are solved 
by the forth-order Runge-Kutta method for 
20 sec with the integration time step of 0.01 msec.
We discard results of initial ten thousand steps to get stable
solutions.
If ISI of spike-train input or output
is about 10 msec, the size of its sample is about 2000.
Although this figure is not sufficiently large for statistics
of ISI data, we hope an essential ingredient will be clarified
in our numerical investigation.

\vspace{0.5cm}
\begin{center}
{\bf III. DETERMINISTIC INPUTS}
\end{center}

\begin{center}
{\bf 3.1 Time-Independent ISI}
\end{center}

\begin{center}
{\bf  A. Pacemaker Neurons }
\end{center}

Let us first consider the HH neuron without the spike-train
input ($I_{\rm p}$ = 0).
The HH neuron is reported to be silent for
$I_{\rm s} = 0$, and to show
the self-excited oscillation 
when $I_s$ exceeds the critical value of 
$I_{\rm ic} = 6.3$ $\mu {\rm A/cm}^2$,
above which $T_{{\rm o} n}$
decreases gradually as $I_s$ is increased.
The dashed curve in Fig.1(d) expresses an example 
of the self-excited oscillation
with the period of $T_{{\rm o} n} = 10.75$ msec
for $I_{\rm s} = 25$ $\mu {\rm A/cm}^2$ and $I_{\rm p} = 0$.

Now we apply the spike-train input to this self-excited 
(pacemaker) neuron.  The input is
given by Eqs.(10)-(16) with the time-independent ISI of 
$T_{{\rm i }n}(t_{{\rm i} n}) = T_{\rm i} = 15$ msec
and $g_{\rm syn} = 0.5 \;{\rm  mS/cm}^2$.
This spike-train input, $U_{\rm i}$, shown in Fig.1(a) yields the
pulse current, $I_{\rm p}$, shown in Fig.1(c), by which
the membrane potential, $V$,  oscillates as depicted
by the solid curve in Fig.1(d).
We plot in Fig.1(b) the time sequence of the spike-train output,
$U_{\rm o}$,  which should be compared with
the input $U_{\rm i}$.
The pulse current, $I_{\rm p}$, has the maximum value of 
$I_{\rm p}^{\rm max} = 14.8$ $\mu {\rm A/cm}^2$ at 
$t =  t_{{\rm i}n} + \tau$  msec.
We notice that the oscillation in $V$ is rather different
from that shown by the dashed curve for $I_{\rm p} = 0$. 
Figure 2 expresses the histogram of the output ISI, showing 
that $\{T_{\rm o}\}$ distributes continuously between 
8.36 to 11.62 msec.
The  mean and root-mean-square (rms)
values of the output ISI are
$\mu_{\rm o} = 10.43$ and $\sigma_{\rm o} = 1.12$ msec.
respectively.
This oscillation is  chaotic  as was pointed
out for the HH neuron receiving sinusoidal inputs
\cite{Aihara84} \cite{Matsumoto84};
the pulse current $I_{\rm p}$ shown in Fig.1(c) is not so different
from the sinusoidal one in a crude sense. 
The chaotic behavior is clearly seen in Fig.3, which 
depicts return maps of input and output ISIs.

When the ISI value of spike-train input, 
$T_{\rm i}$ ($= \mu_{\rm i}$), is changed,
we get an interesting behavior in $\mu_o$ as shown
in Fig.4, where the solid (dashed) curve expresses 
$\mu_{\rm o} \:
(\sigma_{\rm o})$, and filled circles express the distribution of
$\{ T_{{\rm o} n} \}$ for a given $\mu_{\rm i}$.
We note that for $\mu_{\rm i} = 9 - 11$ msec, the period of
the oscillation is forced to be the same;
$T_{{\rm o} m} = T_{\rm i}$, 
leading to the ratio: $k \equiv \mu_{\rm o}/\mu_{\rm i} = 1$.
When $\mu_{\rm i} = 5$ msec, we get $\mu_{o} = 10$ msec and then
$k = 2$.
On the contrary, for $\mu_{\rm i} = 20$ msec, 
we get the two values of $T_{{\rm o} n} = 9$ and 11 msec,
and $\mu_{\rm o} = 10$, the average period of the output 
being a half of the input ($k = 1/2$). 
This is also the case for $\mu_{\rm i} = 21$ and 22 msec.
In the other cases noticed above, the ISI of output
distributes between about 8.5 - 11.5 msec.
We should note that irrespective of $\mu_{\rm i}$, output ISI
is always about 10 msec, which is nearly equal to
$T_{\rm o} = 10.75$ msec,  ISI for
$I_{\rm s} = 25$ and $I_{\rm p} = 0 \: \mu {\rm A/cm}^2$.

\begin{center}
{\bf  B. Silent Neurons }
\end{center}

Next consider the silent neuron with $I_{\rm s} = 0$,
for which the oscillation of the membrane potential 
is induced by applied spike-train inputs.
Figures 5(a)-(d) show the calculated result in which 
the spike-train input is
given by $T_{{\rm i} n} = \mu_{\rm i} = 10$ msec 
and $g_{\rm syn} = 0.5 \: {\rm mS/cm}^2$
without static currents ($I_{\rm s} = 0$).
The applied spike-train inputs shown in Fig.5(a)
create the pulse current with the peaks of 
$I_{\rm p}^{\rm max} = 15.3 \; \mu {\rm A/cm}^2$
as shown in Fig.5(c).
The induced oscillation of the membrane potential, $V$, 
in Fig.5(d) is phase locked with the ratio of $4 : 3$, 
oscillating with a long cycle of
40.00 msec (=11.25 + 12.36 + 16.39) = $4 \: \mu_i$, where
11.25, 12.36 and 16.39 are the values of output ISIs.
The return map of output ISIs is plotted in Fig.6(a).

Figure 7 shows $\mu_o$ and $\sigma_{\rm o}$ as a function of 
$\mu_{\rm i}$.  We notice that $\mu_{\rm o}$ 
agrees with $\mu_{\rm i}$ ($k = 1$)  
for $\mu_{\rm i} $ greater than 12 msec, where the HH neuron
behaves as a simple transmitter with a delay of about 
2.0 msec. 
This is in strong contrast 
with the behavior of the pacemaker neuron discussed
in the preceding sub-section (Fig.4).
On the other hand,
for $\mu_{\rm i}$ less than 11 msec, the behavior of output ISI
is rather complicated. It is easy to see that
$k = 2$ for $\mu_{\rm i}$ = 6, 7 and 8 msec,  and that
$k = 3$ for $\mu_{\rm i}$ = 4 msec.
For $\mu_{\rm i} = 9$, we get $T_{{\rm o} m}$ = 12.06 and 14.96 msec, 
leading to a longer period of $3 \: \mu_i$
= 27.00 ( = 12.06 + 14.96) msec.
For $\mu_{\rm i}$ = 5 msec, we get
$T_{{\rm o} m}$ = 10.94 and 14.06 msec,
which leads to a long period of 5 $\mu_{\rm i}$ = 25 msec,
its return map being shown in Fig.6(b).
Surprisingly, a much longer period of $13 \: \mu_{\rm i}$ is
realized for $\mu_{\rm i} = 11$ msec. 
The rms value of $\sigma_{\rm o}$ has an appreciable value
only around $\mu_{\rm i} = 10$ msec.
 
We have repeated our calculation by changing 
the value of $g_{\rm syn}$.
The calculated ratio, $k = \mu_{\rm o}/\mu_{\rm i}$, is shown as
functions of $g_{\rm syn}$ and $\mu_{\rm i}$ in Fig.8,
where only the integer values of $k$ are shown by
circles ($k=1$), squares($k=2$), 
triangles ($k=3$) and diamond ($k=4$).
Note that non-integer values of $k$ exist between the 
integer values; for example, $k = 4/3$ for 
$g_{\rm syn} = 0.5 \; {\rm mS/cm}^2$ and 
$\mu_{\rm i} = 10$ msec (Fig.7).
We cannot obtain spike-train outputs for small synaptic 
couplings as expected.
When $\mu_{\rm i} = 10$ msec, we get the critical value of
$g_{\rm syn} = 0.11 \; {\rm mS/cm}^2$ below which no outputs
are available.  This coupling yields the pulse current
with the maximum value of 
$I_{\rm p}^{\rm max} = 1.6 \: \mu {\rm A/cm}^2$,
which is  much smaller than the critical dc current
of $I_{\rm ic} = 6.3 \: \mu {\rm A/cm}^2 $ for the self-excited
oscillation with $I_{\rm p} = 0$.
We note that we get $k=1$ for 
the large ISIs with 
fairly strong synaptic couplings.
When we decrease $\mu_{\rm i}$ with keeping
$g_{\rm syn}$ fixed, 
values of $k$ become larger since the HH neuron cannot
respond to inputs with the small ISI because of
its refractory period.
Figure 8 reminds us the result of Guttman, Feldman and
Jakobson \cite{Guttman80} who reported in their Table 1,
the calculated $k$ as functions of the magnitude, $A$,
and the frequency, $f_{\rm i}$, when the sinusoidal input
given by $I_{\rm i} = A \; {\rm sin}(2 \pi\: f_{\rm i}) + I_{\rm b}$ 
is applied to squid giant axons with a bias current, $I_{\rm b}$.
Our result for $\mu_{\rm i} < 10$ msec fairly agrees with 
that of Ref.\cite{Guttman80}.
However, the agreement between the two results is not
good for $\mu_{\rm i} \gg 10$ msec, where 
our input current with the pulse width of 
about $2.45 \: \tau \: \sim 5$ msec 
(Eqs.(15) and (16)) is quite
different from the sinusoidal current adopted in
Ref.\cite{Guttman80}.

As was shown in Fig.4, the pacemaker HH neuron emits the
output ISI of $T_{{\rm o}m} \sim 10$ msec irrespective of 
the value of input ISI, and then it is considered to be
inadequate as a data processor.
Then, in the following sections, we will investigate only
the silent HH neuron with a fixed value of 
$g_{\rm syn} = 0.5$ ${\rm m S/cm}^2$.

\begin{center}
{\bf 3.2 ISI with Sinusoidal Modulation}
\end{center}

In this sub-section we discuss an application of
the spike-train input whose
ISI is modulated by the sinusoidal signal given by
\begin{equation}
T_{{\rm i} n}(t) = d_0 + d_1 \: {\rm sin} ( 2 \pi t/T_p),
\end{equation}
where $T_{\rm p}$ is the period and $d_0$ and $d_1$ are 
coefficients adjusting $\mu_{\rm i}$ and $\sigma_{\rm i}$.

Figure 9(a) and (b) show the time course of input $U_{\rm i}$ 
and output $U_{\rm o}$ for $d_0 = 2 d_1$ = 20 msec
with $T_{p} = 100$ msec.
Because of the introduced sinusoidal modulation,
ISIs at $100 < t < 150$ msec are, for example, larger that those
at $150 < t < 200$ msec.
Figure 9(c) depicts the return map of input
ISIs, which has the egg-shape circle expected
for the sinusoidal signal.
On the other hand, the return map of output ISIs shown in
Fig.9(d) reveals the chaotic behavior.
Results for $d_0 = 2 d_1 = 10$ are plotted in Fig.10(a)-(d),
which show that
although the return map of input ISI has the egg-shape circle,
that of output ISIs is distorted.
The reason of this distortion is explained in Fig.11,
where solid histograms express  input and output ISIs for
$d_0 = 2 d_1 = 10$ 
($\mu_{} = 8.68$, $\sigma_{} = 3.42$)  
and dashed histograms those for
$d_0 = 2 d_1 = 20$ 
($\mu_{i} = 17.54$, $\sigma_{i} = 6.94$), 
with $T_{p} = 100$ msec.
(It is noted that we get $\mu_{i} < d_0$ because the histogram
of the input ISI at $T_{{\rm i}n} < d_0$ has larger magnitudes
than that at $T_{{\rm i} n} > d_0$).
In the case of $d_0 = 2 d_1 = 20$ msec, 
the input and output ISIs distribute almost in the same region
at $11 < T_{{\rm o}m} < 30$ msec.
On the contrary, in the case of $d_0 = 2 d_1 = 10$ msec, 
the output ISIs distribute
at $11.01 < T_{{\rm o}m} < 19.48$ msec while input ISIs are
at $5.00 < T_{{\rm i}n} < 14.96$ msec; 
no output ISIs at $ T_{{\rm o}m} < 11$ msec. This
is due to the refractory period of the HH neuron
and it is the origin of the distortion in the return map
shown in Fig.10(b).
Defining the dimensionless coefficients of variations
for input and output ISIs by
\begin{equation}
c_{\rm v \lambda} = \sigma_{\lambda}/\mu_{\lambda}, 
\:\:\: \:\:\: \:\:\:
\mbox{($\lambda$ = i and o)} 
\end{equation}
we get $c_{\rm vo} = 0.17$ and 0.38 for $d_0 = 2 d_1$ = 10
and 20 msec, respectively; note that $c_{\rm vi} = 0.40$
for both inputs.

Figure 12 shows $\mu_{\rm o}$ and $\sigma_{\rm 0}$ calculated
by changing $\mu_{\rm i}$ with the fixed value of
$c_{\rm vi} = 0.40$.
Solid and dashed curves denote $\mu_{\rm o}$ 
and $\sigma_{\rm o}$,
respectively, and filled circles the distribution
of $\{ T_{{\rm o} m}  \}$ for a given $\mu_{\rm i}$.
We notice that there is no output ISIs with $T_{{\rm o} m}$ less than
about 10 msec, which shows characteristic of
the low-pass filter of the silent HH neuron . 

\vspace{0.5cm}
\begin{center}
{\bf IV. CHAOTIC INPUTS}
\end{center}

\begin{center}
{\bf 4.1 R\"{o}ssler Model}
\end{center}

In this section, we study the spike-train input whose ISI 
is modulated by chaotic signals. 
First we adopt the R\"{o}ssler model, which is given by    
\begin{equation}
d x/dt = - y - z,
\end{equation}
\begin{equation}
d y/dt = x + a \: y,
\end{equation}
\begin{equation}
d z/dt = b \: x - c \: z + x z,
\end{equation}
with $a = 0.36$, $b = 0.4$ and $c = 4.5$ \cite{Rossler74}.
Since ISI has to be  positive and the characteristic 
time scale in the R\"{o}ssler model is different from that
of the HH model, we adopt the variable $x(t)$ which yields
\begin{equation}
T_{{\rm i}n}(t_{{\rm i}n}) 
= d_0 + (d_1/10) \: x(p \: t_{{\rm i}n}),
\end{equation}
with the following two choices of parameters:

\hspace*{3cm}
$d_0 = d_1 = 10$ msec and $p = 1/10 \;\;\;\;$  (case R1),

\hspace*{3cm}
$d_0 = d_1 = 20$ msec and $ p = 1/20 \;\;\;\;$ (case R2). 

Figure 13(a) and (b) show the time course of input and output 
spike trains for the case R1.
The return map of input ISI depicted in Fig.13(c)
shows a shape characteristic for chaotic signals.
On the other hand, the return map of output ISIs is rather strange
with no traces at $T_{\rm o} < 10$ msec.
This is due to the low-pass filter behavior of the silent 
HH neuron, as shown by solid histograms in Fig.14(a) and (b);
output ISIs distribute at $11.11 < T_{{\rm o}m} < 25.15$ msec
($\mu_{{\rm o}m} = 13.43, \sigma_{\rm o} = 2.44$ msec) whereas
input ISIs distribute at $5.06 < T_{{\rm i}n} < 16.56$ msec
($\mu_{\rm i} = 9.53, \sigma_{\rm i} = 2.69$ msec).

Return maps for the case R2 are shown in Fig.15, 
in which both return maps are almost the same.  This is
because input and output ISIs locate almost in the same region of
$10  < T_{{\rm i}n}, \; T_{{\rm o}m} < 30$ msec, as shown by
dashed histograms in Fig.14(a) and (b). 

Next we investigate the nature of the correlation in the
ISI sequences.
This is made by employing the surrogate data method
applied to ISI data \cite{Theiler92}.
We adopt the shuffled surrogate as a simple method to get
surrogate data. The distributions of ISIs
of shuffled surrogate inputs are exactly the same as those of 
original ISI data although surrogate data have no correlation
between successive ISI values.

The time course of the membrane potentials
for the surrogate data is ostensibly quite
similar to that for the original chaotic input (not shown).
The solid (dashed) histogram in  Fig.14(c) shows the distribution
of output ISIs of surrogate data generated from the R\"{o}ssler
model for the case R1 (R2).  The results of the surrogate data
are similar to those for the corresponding original data.
Return maps of the input and output ISI of the surrogate data,
depicted in Fig.15(c) and (d), show the characteristics
of random signals.

\begin{center}
{\bf 4.2 Lorentz Model}
\end{center}

The similar calculation is made with the use of the 
Lorentz model, which is given by
\begin{equation}
d x/dt = d \: (y - x),
\end{equation}
\begin{equation}
d y/dt = e \: x - y - x \: z,
\end{equation}
\begin{equation}
d z/dt = - f \: z + x \: y,
\end{equation}
with $d = 10$, $e = 28$ and  $f = 8/3$ \cite{Lorentz63}.
We employ the variable $z(t)$, with which
the input ISI is given by
\begin{equation}
T_{{\rm i}n}(t_n) 
= d_0 + (d_1/25) \;\; (z(p t_{{\rm i}n}) - 25),
\end{equation}
where $d_0 = d_1 = 20$ and $p = 1/100$. 

Figures 16(a)-(d) show return maps of ISI data of original chaotic data
and its surrogate.
Return maps of output ISIs for the chaotic and surrogate data
shown in Fig.16(b) and (d) have no traces at $T_{{\rm o}m} > 10 $ msec 
because of the low-pass filter character of the HH neuron.

Sauer \cite{Sauer94}, and  Racicot and Longtin \cite{Racicot97}
studied the response of the IF model to the input whose 
amplitudes are modulated by chaotic signals.
It was shown that when the mean firing rate is high,
the relationship between input and output is high, which leads to
the  high nonlinear predicability.
Our calculations have shown that
when the mean firing rate is too high 
({\it i.e.} input ISIs are too short such as $T_{{\rm i}n} < 10$ msec),
the information is lost because the HH neuron behaves as the low-pass
filter due to its refractory period, which is not included in the
IF model.

\vspace{0.5cm}
\begin{center}
{\bf V. STOCHASTIC INPUTS}
\end{center}

The ISIs of spike-train input, $T_{{\rm i} n}$, 
in Eq.(13) are assumed to be
independent random variables with the Gamma probability
density function given by
\begin{equation}
P(T) = s^r \;\; T^{r - 1} \;\; e^{- s T}/\; \Gamma(r)
\end{equation}
for which we get $\mu_{\rm i} =  r/s$,
$\sigma_{\rm i} = \sqrt{ r}/s$ and
$c_{vi} = 1/\sqrt{r}$,
$\Gamma \;(r)$ being the gamma function.
It is noted that in the limit of $r = 1$,
Eq.(29) reduces to the Poisson distribution 
($c_{\rm vi}$ = 1) and
that in the limits of $r \rightarrow \infty$ 
and $s \rightarrow \infty$ with keeping 
$\mu_{\rm i} = r/s $ fixed, Eq.(29) reduces to
$P(T) = \delta(T - \mu_{\rm i}) $, the constant ISI
with $\mu_{\rm i} = T$ and $c_{\rm vi} = 0$.

The spike-train input created by the Gamma-distribution 
generator is applied to our neural system.
Calculations are performed by changing $\mu_{\rm i}$
by keeping the value of 
$c_{\rm vi}$ fixed.
Note that because the number of our sample of
input ISI is not sufficiently large, the obtained
$c_{\rm vi}$ fluctuates around the intended values.
Solid histograms in Fig.17(a) and (b) show
the result for $c_{\rm vi}$ = 0.40, $\mu_{\rm i} = 10$ msec, 
$c_{\rm vo}$ = 0.25 and $\mu_{\rm o}$ = 14.84 msec 
while dashed histograms for 
the result for $c_{\rm vi}$ = 0.40, $\mu_{\rm i} = 20$ msec, 
$c_{\rm vo}$ = 0.36 and $\mu_{\rm o}$ = 21.11 msec. 

Solid and dashed curves in Fig.18(a) denote $\mu_{\rm o}$
and $\sigma_{\rm o}$, respectively, for
$c_{\rm vi} = 0.4$.
We note that as increasing $\mu_{\rm i}$, 
$\mu_{\rm o}$ increases and approaches the dotted line expressing
$\mu_{\rm o} = \mu_{\rm i}$. 
This is similar to the case of $c_{\rm vi} = 0$ shown in Fig.7,
where $\mu_{\rm o} = \mu_{\rm i}$ at $\mu_{\rm i } \simg 10$ msec. 
On the contrary, the dependence of 
$\mu_{\rm o} $ on $\mu_{\rm i}$ for
the case of $c_{\rm vi} = 1.0$ shown in Fig.18(b), is rather different 
from the cases of $c_{\rm vi}$ = 0 and 0.4.
We get $\mu_{\rm o} \sim (\mu_{\rm i} + 10)$ msec 
at $\mu_{\rm o} < 100$ msec
and it 
deviates from the dotted line showing 
$\mu_{\rm o} = \mu_{\rm i}$.
These calculations depicted in Figs.18(a) and (b) 
clearly show that 
$\mu_{\rm o}$ depends not only on $\mu_{\rm i}$ 
but also on $\sigma_{\rm i}$ ($c_{\rm vi}$).
Although both $\mu_{\rm o}$ and $\sigma_{\rm o}$ increase
as the value of $\mu_{\rm i}$ is increased, 
the increase in the latter is more
significant than that in the former, which
yields an increase in $c_{\rm vo}$, as shown by
the thin-solid curves in Figs.18(a) and (b).
We note that $c_{\rm vo}$ approaches the values of $c_{\rm vi}$
as increasing $\mu_{\rm i}$, the related discussion
being given in Sec.VI.

We have performed the calculation also using input ISI with
random, uniform distribution.
Obtained results are similar to those for
the Gamma distribution, as far as the adopted values
of $c_{\rm vi}$ are the same (not shown).

\vspace{0.5cm}
\begin{center}
{\bf VI. CONCLUSION AND DISCUSSION}
\end{center}

We have investigated the responses of the HH neurons,
by applying the various types of spike-train inputs
whose ISI is modulated by deterministic, chaotic and
stochastic signals.
The obtained results are summarized as follows:

\noindent
(1) Output ISIs of the pacemaker HH neuron against the
time-independent input ISI are always 
$T_{{\rm o}m} \sim 10$ msec
irrespective of the value of the input ISI (Fig.4).

\noindent
(2) Output ISIs of the silent HH neurons for the constant
ISI with $T_{{\rm i}n} > 10$ msec yield output ISI 
with $T_{{\rm o}m} = T_{{\rm i}n}$  whereas
for ISI with $T_{{\rm i}n} < 10$ msec,  the HH neuron generally
emits multiple numbers of output ISIs (Fig.7).

\noindent
(3) For the input ISI modulated by sinusoidal, chaotic and
stochastic signals, the silent HH neuron behaves
as a low-pass filter because of its refractory period,
yielding output ISI with $T_{{\rm o}m} > 10$ msec.


\noindent
(4) Output ISIs generally depend not only
on the mean of the input ISI but also on their
fluctuations: the HH neuron is not a simple integrator.

\noindent
(5) The analysis on the histograms of input and output ISIs
cannot distinguish the responses to the deterministic,
chaotic and stochastic signals.

\noindent
(6) The distinction can be made by an analysis of the time 
correlation of the ISI data, for example, by
plotting their return maps.

Softky and Koch \cite{Softky92} have reported a large 
coefficient of variation ($c_{\rm vo} \; = 0.5 \sim 1.0)$ 
for spike trains of non-bursting cortical neurons 
in visual V1 and MT of monkeys in strong contrast with
a small $c_{\rm vo}\; ( = 0.05 \sim 0.1)$ in motor neurons
\cite{Calvin68}.
In order to explain the large $c_{\rm vo}$, several hypotheses
have been proposed; a balance between excitatory and
inhibitory inputs \cite{Shadlen94},
the high physiological gain in the 
$f_{\rm o}-I_{\rm i}$ plot \cite{Troyer98}, 
correlation fluctuations in recurrent 
networks \cite{Usher94},
and the active dendrite conductance \cite{Softky95}.
By using the IF model,
Feng and Brown \cite{Feng98} have shown that
there are three kinds of behaviors of 
$c_{\rm vo}$ depending on
the distribution of input ISIs;

\noindent
(a) $c_{\rm vo}$ tends to decrease for the Gaussian,
uniform or truncated distribution of ISIs,

\noindent
(b) $c_{\rm vo}$ remains constant for the exponentially 
distributed ISIs, and

\noindent
(c) $c_{\rm vo}$ diverges to infinity when ISIs follow
the Pareto distribution which has a slow-decreasing 
tail of $T^{- \alpha}$ ($\alpha > 0$) at large $T$.

\noindent
The case (a) was previously discussed by
Marsalek, Koch and Maunsell \cite{Marsalek97}.

Figure 19 shows the dependence of $c_{\rm vo}$ on $c_{\rm vi}$
for various types of our input ISIs having been
reported in previous sections.
Inverted triangles denote the results for the constant ISIs
($\S$3.1B), open marks the results for input ISIs with
sinusoidal modulation ($\S$ IV), and
filled circles, triangles and squares
results of the stochastic modulation ($\S$ V).
Our calculations show the followings:

\noindent
(i) The constant ISI with a vanishing $c_{\rm vi}$
yields the finite $c_{\rm vo}\: (< 0.2)$,
{\it i.e. } $c_{\rm vo} \simg c_{\rm vi}$, 

\noindent
(ii) the finite-width distribution of ISIs with the sinusoidal
modulation leads to   $c_{\rm vo} \siml c_{\rm vi}$, and 

\noindent
(iii) the exponential, Gamma distribution of input ISIs
yield the result which is ostensibly similar to that in 
the item (ii). 

\noindent
Although the item (ii) is in agreement with the result for 
the above-mentioned case (a) 
discussed in Refs.\cite{Feng98} and \cite{Marsalek97},
the items (i) and (iii) disagree with the results of the
cases of (a) and (b), respectively.
This difference is expected to arise from the fact
that the response of the HH neuron with the refractory period
is different from that of the IF neuron without it
\cite{Gutkin98}.

Finally we want to discuss the transient response of the HH neuron
to the cluster of spike-train inputs.
Figure 20(a), (b), and (c) show the results
for $T_{{\rm i}n}$ = 5, 10 and 20 msec, respectively.
In Fig.20(c), for example, the upper (lower) panel
of C1, C2, C3 and C4 express the time courses of input 
(output) spike trains for inputs of two, three, four and 
five impulses, respectively, with $T_{{\rm i}n}$ = 20 msec.
In this case the ISI of output pulses is the same as that of 
input pulses, 
$T_{{\rm o}m} = T_{{\rm i}n}$, because the HH neuron
behaves as a linear transmitter 
for inputs with ISI of $T_{{\rm i}n} \simg 10$ msec.
On the contrary, its behaviors of output ISI data 
for inputs with
$T_{{\rm i} n}$ = 5 and 10 msec are much complicated.
Fig.20(a) shows that
output ISIs for $T_{{\rm i}n}$ = 5 msec are 11.39 and 11.87 msec, 
which should be compared with $T_{{\rm o}m}$ = 10.94 and 14.06 msec
for the sequence of the spike trains with
constant ISI of 5 msec discussed in sec. 3.1B.
For the case of $T_{{\rm i}n}$ = 10 msec shown in Fig.20(b), 
we get $T_{{\rm o}m}$ = 11.44, 11.80 and 17.11
msec whereas the sequence of the constant ISI of
10 msec leads to $T_{{\rm o}m}$ = 11.25, 12.36 
and 16.39 msec (Fig.5).
It is noted that  both inputs with three (A2) 
and four impulses (A3) yield the same output of two impulses.
Similarly, inputs with three (B2) and four impulses (B3)
lead to outputs with three impulses.
We should note in all the cases shown in Fig.20(a)-(c) that 
the first output pulse is rather 
quickly emitted with a delay of 2.1 msec 
after the first input pulse of clusters is applied to the HH neuron.
This fast transient response may be relevant to a quick 
passage of information reported by Thorpe, Eize and Marlot 
\cite{Thorpe96} and by Rolls and Tovee \cite{Rolls94}.

\section*{Acknowledgements}
This work is partly supported by
a Grant-in-Aid for Scientific Research from the Japanese 
Ministry of Education, Science and Culture.



\begin{figure}
\caption{
Responses of the pacemaker HH neuron to the time-independent
input ISI ($I_{\rm p}$ with $T_{\rm i}{\rm} = 15$ msec and 
$I_{\rm s} = 25 \: \mu {\rm A/cm}^2$;
time sequences of (a)  input $U_{\rm i}$, 
(b) output $U_{\rm o}$,
(c) pulse current $I_{\rm p}$, and
(d) membrane potential $V$,
dashed curves in (b) and (d) denoting the result with
$I_{\rm p} = 0$.
}
\label{fig1}
\end{figure}

\begin{figure}
\caption{
Histograms of (a) the time-independent 
input ISI ($T_{{\rm}i}$ = 15 msec) and 
(b) output ISI of the pacemaker
neuron (see Fig. 1).
}
\label{fig2}
\end{figure}

\begin{figure}
\caption{
Return maps of (a) time-independent input ISI with $T_{\rm i} = 15$ msec
and (b) output ISI of the pacemaker neuron (Fig.1).
}
\label{fig3}
\end{figure}

\begin{figure}
\caption{
Mean ($\mu_{\rm o}$, solid curve) and 
rms ($\sigma_{\rm o}$, dashed curve)
values of output ISI of pacemaker neurons 
($I_{\rm s} = 25 \: \mu \: {\rm A/cm}^2$)
against the mean value ($\mu_{\rm i}$)
of time-independent input ISI.
Filled circles denote the distribution of output ISIs
for a given $\mu_{\rm i}$,
dotted curves denoting
$k \equiv \mu_{\rm o}/\mu_{\rm i}$.
}
\label{fig4}
\end{figure}

\begin{figure}
\caption{
Response of the silent HH neuron to 
the time-independent input ISI
($I_{\rm p}$ with $T_{\rm i} = 10$ msec and 
$I_{\rm s} = 0)$;
time sequences of (a) the input $U_{\rm i}$, 
(b) output $U_{\rm o}$,
(c) pulse current $I_{\rm p}$, and
(d) membrane potential $V$.
}
\label{fig5}
\end{figure}

\begin{figure}
\caption{
Return maps of output ISIs for the time-independent input
with (a) $T_{\rm i}$ = 10 and (b) 5 msec (see Fig.5).
}
\label{fig6}
\end{figure}

\begin{figure}
\caption{
Mean ($\mu_{\rm o}$, solid curve) and 
rms ($\sigma_{\rm o}$, dashed curve)
values of output ISI of silent neurons 
against the mean value ($\mu_{\rm i}$)
of time-independent input ISI.
Filled circles denote the distribution of output ISIs
for a given $\mu_{\rm i}$,
dotted curves denoting $k \equiv \mu_{\rm o}/\mu_{\rm i}$.
}
\label{fig7}
\end{figure}

\begin{figure}
\caption{
The calculated ratio, $k=\mu_{\rm o}/\mu_{\rm i}$, as functions of
$\mu_{\rm i}$ and $g_{\rm syn}$ for the time-independent ISI input to 
silent neurons: only integer $k$'s are
shown by circles ($k=1$), squares ($k=2$), 
triangles ($k=3$) and diamond ($k=4$),
the cross denoting no outputs.
}
\label{fig8}
\end{figure}

\begin{figure}
\caption{
Time courses of (a) input $U_{\rm i}$ and (b) output $U_{\rm o}$,
and return maps of (c) input and (d) output ISIs for
the sinusoidal modulation for $d_0 = 2 d_1 = 20 $ msec (Eq.19).
}
\label{fig9}
\end{figure}

\begin{figure}
\caption{
Time courses of (a) input $U_{\rm i}$ and (b) output $U_{\rm o}$,
and return maps of (c) input and (d) output ISIs for
the sinusoidal modulation for $d_0 = 2 d_1 = 10 $ msec (Eq.19).
}
\label{fig10}
\end{figure}

\begin{figure}
\caption{
Histograms of (a) input ISI with sinusoidal modulation
and (b) output ISI.  Solid (dashed) curves are for 
$d_0 = 2 d_1 $ = 10 (20) msec.
}
\label{fig11}
\end{figure}

\begin{figure}
\caption{
Mean ($\mu_{\rm o}$, solid curve) and 
rms ($\sigma_{\rm o}$, dashed curve)
values of output ISI against the mean value ($\mu_{\rm i}$)
of input ISI with sinusoidal modulation ($c_{\rm vi}$ = 0.40).
Filled circles denote the distribution of output ISIs
for a given $\mu_{\rm i}$,
dotted curves denoting $k \equiv \mu_{\rm o}/\mu_{\rm i}$.
}
\label{fig12}
\end{figure}

\begin{figure}
\caption{
Time courses of (a) input $U_{\rm i}$ and (b) output $U_{\rm o}$,
and return maps of (c) input and (d) output ISIs for
chaotic inputs generated by the R\"{o}ssler model (case R1).
}
\label{fig13}
\end{figure}

\begin{figure}
\caption{
Histograms of (a) input ISI and (b) output ISI
for the chaotic input generated
by the R\"{o}ssler model,
and (c) output ISI for its surrogate,
solid (dashed) curves being for case R1 (R2).
}
\label{fig14}
\end{figure}

\begin{figure}
\caption{
Return maps of (a) input ISI and (b) output ISI
for the chaotic input generated
by the R\"{o}ssler model (case R2); (c) and (d) are
corresponding return maps  of its surrogate data.
}
\label{fig15}
\end{figure}

\begin{figure}
\caption{
Return maps of (a) input ISI and (b) output ISI
for the chaotic input generated
by the Lorentz model; (c) and (d) are
corresponding return maps  of its surrogate data.
}
\label{fig16}
\end{figure}

\begin{figure}
\caption{
Histograms of (a) input ISI and (b) output ISI
for spike-train inputs with the Gamma distribution;
solid (dashed) curves for input ISIs of
$\mu_{\rm i}$ = 10 (20) msec with
$c_{\rm vi}$ = 0.40. 
}
\label{fig17}
\end{figure}

\begin{figure}
\caption{
Mean ($\mu_{\rm o}$, solid curves),
rms ($\sigma_{\rm o}$, dashed curves) and
$c_{vo}$ (thin solid curves)
of output ISIs against the mean value ($\mu_{\rm i}$)
of input ISI for Gamma distribution with
(a) $c_{\rm vi}$ = 0.4 and
(b) $c_{\rm vi}$ = 1.0.
The dotted curves denoting 
$\mu_{\rm o} = \mu_{\rm i}$
are plotted for a guide of the eye
}
\label{fig18}
\end{figure}

\begin{figure}
\caption{
$c_{\rm vo}$ against $c_{\rm vi}$
for inputs with time-independent ISI (inverted triangles,
$c_{\rm vi} = 0$), with sinusoidal modulation
(open diamonds, squares and triangles for
$c_{\rm vi}$ = 0.22, 0.40 and 0.82, respectively),
and with random Gamma distribution 
(filled squares, triangles
and inverted triangles for 
$c_{\rm vi}$ = 0.40, 0.75 and  1.03, respectively).
The dotted curve expressing 
$c_{\rm vo} = c_{\rm vi}$ is plotted
for a guide of the eye.
}
\label{fig19}
\end{figure}

\begin{figure}
\caption{
Time courses of spike-train inputs and outputs;
input ISIs are (a) $T_{i}$ = 5, (b) 10 and (c) 20 msec, and
upper (lower) panel of each figure shows inputs (outputs).
}
\label{fig20}
\end{figure}

\end{document}